\documentclass[a4paper]{llncs}

\usepackage[pdftex]{graphicx}
\usepackage{url}
\usepackage{placeins}

\newcommand{\fref}[1]{\figurename~\ref{#1}}
\newcommand{\sref}[1]{Sect.~\ref{#1}}
\newcommand{\eref}[1]{Example~\ref{#1}}
\newcommand{\dref}[1]{Def.~\ref{#1}}

\newcommand{\flora}{$\mathcal{F}LORA$-2}
\newcommand{\dlpm}{Datalog$^\pm$}
\newcommand{\code}[1]{{\small\ttfamily{#1}}}
\newcommand{\defeq}{\stackrel{\textup{\tiny def}}{=}}

\begin{document}
  \sloppy
\mainmatter

\title{Rule Module Inheritance\\with Modification Restrictions}

\titlerunning{Rule Module Inheritance}
\author{Felix Burgstaller\inst{1} \and Bernd Neumayr\inst{1} \and Emanuel Sallinger\inst{2} \and Michael Schrefl\inst{1}}

\institute{Johannes Kepler University Linz, Linz, Austria \email{\{firstname.surname\}@jku.at} \and
University of Oxford, Oxford, United Kingdom \email{emanuel.sallinger@cs.ox.ac.uk}}

\maketitle              

\begin{abstract}
  Adapting rule sets to different settings, yet avoiding uncontrolled proliferation of variations, is a key challenge of rule management.
  One fundamental concept to foster reuse and simplify adaptation is inheritance. Building on rule modules, i.e., rule sets with input and output schema, we formally define inheritance of rule modules by incremental modification in single inheritance hierarchies. To avoid uncontrolled proliferation of modifications, we introduce formal modification restrictions which flexibly regulate the degree to which a child module may be modified in comparison to its parent.
  As concrete rule language, we employ \dlpm~which can be regarded a common logical core of many rule languages. We evaluate the approach by a proof-of-concept prototype.
\end{abstract}

\section{Introduction}
In data- and knowledge-intensive systems it is good practice to separate explicit knowledge, elicited from domain experts and translated into declarative expressions by rule developers, from application code developed by application developers.
Rule-based knowledge representation and reasoning build the core of systems for business rule engines~\cite{OMG2017}, web data extraction~\cite{Furche2014}, data wrangling~\cite{Konstantinou2017}, knowledge graph management~\cite{Bellomarini2017a}, and information tailoring~\cite{Burgstaller2015}. With the increasing number and complexity of rules, their maintenance and their adaptation to different settings and contexts become key challenges of rule developers.
In this paper we present an approach employing rule modules and inheritance to cope with these challenges. In the following we sketch challenges and our approach along a self-contained business rule example which is used throughout the paper. Subsequently, we zoom out to give the big picture of potential application areas where the presented approach may serve as a central building block.

\subsection{Challenges and Approach}
In this paper, we present an approach which builds on \emph{rule modules} -- i.e, rule sets with interfaces describing the schema of input and output data -- to provide a clear separation and interfaces between rule sets and data-intensive applications. These interfaces shield application developers from the intricacies of rule sets as well as rule developers from application code and clarify which knowledge and data schemata are internal to the module and which interface either as input or output. --- For example, a bank clerk is responsible for processing a set of mortgage applications, i.e., assessing the applications' credit worthiness in order of their priority. Domain knowledge regarding assessment and prioritization of mortgage applications is encoded in a rule module \code{MortgageApps}. This rule module is employed by a data-intensive application collecting and managing mortgage applications and their assessment. Rule module \code{MortgageApps} takes as input a set of mortgage applications each described by the mortgage value and the estimated values of real estate securities. As output it produces a preliminary assessment of the credit worthiness, either \code{good} or \code{bad}, of each application together with a prioritization of the applications for detailed assessment. Any issues with applications, e.g., having a mortgage value below the specified minimum loan value, are also output.

Organizing business rules into rule modules entails the danger of \emph{redundancy}: One and the same rule may be relevant in different settings and thus introduced and maintained separately in different modules. This duplicates human effort in developing and maintaining modules and makes it difficult to keep rules synchronized across modules. --- For example, our bank decides to offer private loans as well, creating a rule module \code{PrivateLoanApps}. Some rules, such as the minimum loan value, apply in module \code{MortgageApps} as well as module \code{PrivateLoanApps}. Thus, any changes to this rule need to be performed in both rule modules.

In this paper we introduce \emph{inheritance} of parent modules to child modules by incremental modification as one way to mitigate these problems without sacrificing flexibility. In child modules, rule developers may introduce additional rules, remove inherited rules, as well as extend and/or reduce the input and output interfaces. This reduces redundancy and thus should ease maintenance of modules adapted to different business settings. --- For example, extracting common rules, such as the minimum loan value rule, into a parent module \code{LoanApps} we can remove redundancy. Any changes to a common rule are made in the parent module and by inheritance propagated to all child modules. Moreover, since we allow modifications to inherited rules, we can define default rules for loan application assessment and ranking in module \code{LoanApps}.

Allowing arbitrary modifications in child rule modules, however, would undermine the benefits of inheritance and would pave the way for \emph{uncontrolled proliferation of variations}: a rule or application developer trying to get an overall picture of the interfaces and the behavior, i.e., the derived knowledge, encoded in a hierarchy of rule modules would still have to inspect all modules. Furthermore, undesired changes to inherited rules and interfaces would be possible. --- For example, consider our minimum loan value rule in module \code{LoanApps}. Since arbitrary modifications are allowed we can simply remove this rule in child modules; any problematic loan applications would not be output anymore.

In this paper, we introduce a set of \emph{modification restrictions} to flexibly regulate the degree to which a module's interfaces and behavior may be modified in comparison to its parent module's interfaces and behavior. Treating rule modules as black boxes, modification restrictions set boundaries within which a module may be modified. Thus, it should be sufficient to inspect the root module to get an abstract overview of the behavior implemented in a hierarchy of modules.

Structural restrictions restrain allowed modifications to the input and output schema of a module. --- For example, module \code{LoanApps} specifies the basic input for loan application assessments. Child module \code{MortgageApps} requires further inputs like real estates provided as securities. Thus, we define \code{LoanApps}' inputs as non-omitable. Moreover, module \code{MortgageApps} fixes the output for all mortgage application modules. Thus, we define its output as non-extensible.

Behavioral restrictions constrain modifications changing the output at instance level (for a particular output schema element a behavioral modification may lead to different instances in the output). --- For example, we want to prohibit child modules from deriving a subset of minimum loan value issues compared to module \code{loanApps}. Thus, we employ restriction non-shrinkable. Similarly, child modules must not have weaker requirements for good credit worthiness, i.e., must not derive good credit worthiness for a superset of loan applications compared to module \code{loanApps}. Thus, we employ restriction non-growable.

\subsection{Potential Application Areas}
Potential application areas, besides business rules as shown in the examples, are rule-based systems for information tailoring~\cite{Burgstaller2015}, web data extraction~\cite{Furche2014}, and knowledge graph management~\cite{Bellomarini2017a}, where rule modules encode knowledge for tasks such as data extraction, transformation, cleansing, and filtering and need to be adapted to different settings.

For example, in the DIADEM system for web data extraction~\cite{Furche2014}, multiple rule modules, 
each responsible for a particular task such as web form understanding or form filling, are dynamically orchestrated in networks where one module's output is another module's input. Some of the encoded knowledge necessary for web form understanding is generic to 'all' web sites, while other encoded knowledge is specific to domains such as 'real estate' websites. Inheritance from the rule module for generic web form understanding to a rule module for web form understanding for 'real estate' websites should help to reduce redundancy and thus ease maintenance. Structural modification restrictions can be used to ensure that child rule modules remain orchestrable, similar to co- and contravariance in object-orientation which can be used to ensure type-safety and substitutability. Behavioral modification restrictions should help to keep the overall rule base understandable and thus maintainable, e.g., looking at a parent module and its behavioral restrictions gives an overview of possible behavior in child modules.

Regarding Web developments, e.g., Internet of Things, Semantic Web, or Smart Things, rules become a vital and integral part \cite{Weigand2012}. Due to the number of rules and their context-dependent applicability, efficient rule management is essential. Rule modules and module inheritance are means to manage such large rule sets and can be extended to manage contexts of application (c.f. \cite{Burgstaller2017c}).

We expect our approach to be applicable and beneficial in these areas -- an evaluation is yet to be performed. In this paper we focus on rule modules, their inheritance, and modification restrictions independent of specific applications.

\subsection{Contributions and Overview}
Currently, work on rule inheritance and related work on contextualized knowledge representation is fragmented and there exists no approach for inheritance of rule modules where modifications can be flexibly restricted. --- In this paper we introduce an overall approach to rule module inheritance and specifically make the following contributions:
\begin{itemize}
     \item formal definition of (1) downward rule and interface inheritance in single inheritance hierarchies of modules, and (2) modification restrictions and inheritance of modification restrictions
     \item discussion of conformance checks for detecting violations of defined structural and behavioral modification restrictions
     \item proof-of-concept prototype implementing formal definitions in Datalog, publicly available for further experimentation
\end{itemize}

The remainder of this paper is structured as follows: \sref{sec:BRModules} presents our \dlpm-based rule language and  rule modules. In \sref{sec:inheritance} we present our basic inheritance mechanism. \sref{sec:restrictions} introduces modification restrictions and delineates inheritance of modification restrictions. In \sref{sec:eval} we present a proof-of-concept prototype. \sref{sec:related} discusses related work. \sref{sec:conclusion} concludes the paper.

\section{Rule Modules}\label{sec:BRModules}
First, we discuss expressing rules with \dlpm~where a rule is constructed from predicates and operators like logical conjunction. Subsequently, we introduce rule modules and discuss their structure and behavior.

\subsection{Underlying Rule Language}
This section delineates our notion of rules based on a formal language. In order to focus on rule module inheritance and modification restrictions, the underlying rule language and data model should be simple, i.e., have few constructs and operators, to avoid unnecessary complexity. A fitting family of formal languages is \dlpm~\cite{Cali2012} which has clearly defined formal semantics and employs a relational data model.

\dlpm~extends plain Datalog by existentially quantified variables in rule heads, negative constraints, and equality-generating dependencies while restricting the language so as to achieve decidability and, more particularly, good performance. Vadalog~\cite{Bellomarini2018} is a practical implementation of \dlpm~that adds many features needed in commercial use, including a wide range of built-ins. Thus, \dlpm/Vadalog is quite expressive, e.g., encompassing full Datalog with no restriction on recursion and SPARQL under the OWL 2 QL entailment regime, while still being efficient. \dlpm /Vadalog is a versatile rule language family also employed for big data wrangling~\cite{Konstantinou2017} and knowledge graph management~\cite{Bellomarini2017a}.

A plain Datalog \emph{rule} comprises a body (premise) and a head (conclusion) where the conclusion is derived if the premise holds. Both conclusion and premise are conjunctions of atoms (i.e., predicates with arguments). \dlpm~allows use of existentially quantified variables in conclusions enabling value creation, truth value \code{false} in conclusions (negative constraints), and equality-generating dependencies, e.g., \code{Y=Z:- r1(X,Y), r2(X,Z)}.

\begin{definition}[Rule Structure]\label{def:rule}
  Predicates (a.k.a. relations) are taken from a universe of predicates $P$.
  Rules are taken from a universe of rules $R$.
  A rule $r \in R$ has body predicates $B_r \subseteq P$ and head predicates $H_r \subseteq P$.
\end{definition}

\begin{example}\label{ex:rule}
  Our bank deems mortgage loans of less than 10,000 Euro as not worth the organizational costs (rule \code{R0}). We translate this natural language rule to~\dlpm: \code{lowLValue(X,V) :- lValue(X,V), V < 10000.} The mortgage value applied for is represented by \code{lValue/2} (predicate \code{lValue} with arity two) relating loan applications with loan values while \code{lowLValue/2} annotates mortgage applications below the defined mortgage value threshold. Rule \code{R0} has body predicates $B_{{\tt R0}} = \{$\code{lValue/2}$\}$ and head predicates $H_{{\tt R0}} = \{$\code{lowLValue/2}$\}$.
\end{example}

\subsection{Module Structure and Behavior}
A collection of rules and facts (i.e., rules without variables and premise true) in Datalog is called a Datalog program. In the following we extend Datalog programs with input schema and output schema to rule modules, where the Datalog program itself is the implementation of the rule module.

We derive \emph{structural} aspects of rule modules from Datalog and Vadalog. Datalog splits the predicates of a program into extensional database (EDB) and intensional database (IDB). EDB contains predicates asserted in the knowledge base, IDB predicates defined by rules. A similar distinction is made in DMN's rule representation~\cite{OMG2016}. Vadalog~\cite{Bellomarini2018} extends this idea to \emph{inputs} provided by external sources (input predicates) and derived predicates \emph{output} to external sinks (output predicates). These two sets are disjoint. Predicates derived and potentially used in rule bodies but not exported are auxiliary predicates.

\begin{definition}[Rule Modules]
  Rule modules are taken from a universe of rule modules $M$.
  A rule module $m \in M$ is defined by a set of rules $R_m$, a set of input predicates $I_m \subseteq P$, and a set of output predicates $O_m \subseteq P$.
  The sets of input and output predicates are disjoint.
  The predicates of a module $m$ ($P_m$) are the union of its rule head and body, input, and output predicates.
\end{definition}

We now discuss the development of a rule module from an organizational perspective: Domain experts elicit and determine rules, often in natural language, and organize them into rule sets, e.g., a rule set may cover a specific business case. Furthermore, domain experts determine the necessary data input and derived output for the elicited set of rules. Once the domain experts consider a rule set and its inputs and outputs complete, rule developers translate the elicited rules and interfaces into a rule module with a \dlpm~program as implementation.

\begin{example}\label{ex:module}
  We want to elicit for the process of assessing mortgage loan applications all relevant rules. In addition to rule \code{R0} (see \eref{ex:rule}) we identified: Credit worthiness of a mortgage application is deemed good (predicate \code{cwGood/1}) if the value of provided properties exceeds 80 \% of the mortgage's value (\code{R1}). In all other cases credit worthiness is deemed bad (\code{R2} -- output \code{cwBad/1}). Loan applications of higher loan values have priority over those with lower loan values (\code{R3} -- \code{priorityOver/2}). For each provided property, its value needs to be calculated (\code{sValue/2}) and associated with the loan application (\code{R4} -- \code{property/1, properties/2}). Associated properties are securities (\code{R5} -- \code{securities/2, security/1}). Properties of a value below 30,000 Euro are problematic as securities (\code{R6} -- \code{lowPropValue/2}). The predicates derived by rules \code{R0-R6} form module \code{MortgageApps}'s output predicates $O_{\tt MortgageApps}$.

  From these natural language rules our domain experts derive the necessary input. The rules employ information given with each mortgage application (\code{loan/1}). Such applications need to state the intended mortgage value (\code{lValue/2}), the duration (\code{duration/2}), the applying customer (\code{customer/2}), and any real estate properties which may be used as securities (\code{mProperty/2}) as well as their value (\code{pValue/2}). These predicates form the input predicates $I_{\tt MortgageApps}$. Furthermore, properties may be hierarchically organized (\code{hasPart/2}), e.g., a property may consist of an area containing buildings. A visual summary including the \dlpm~representation is given in \fref{fig:MortgageApps}.

      \begin{figure}[t!]
      	\centering
      	\includegraphics[width=.95\textwidth]{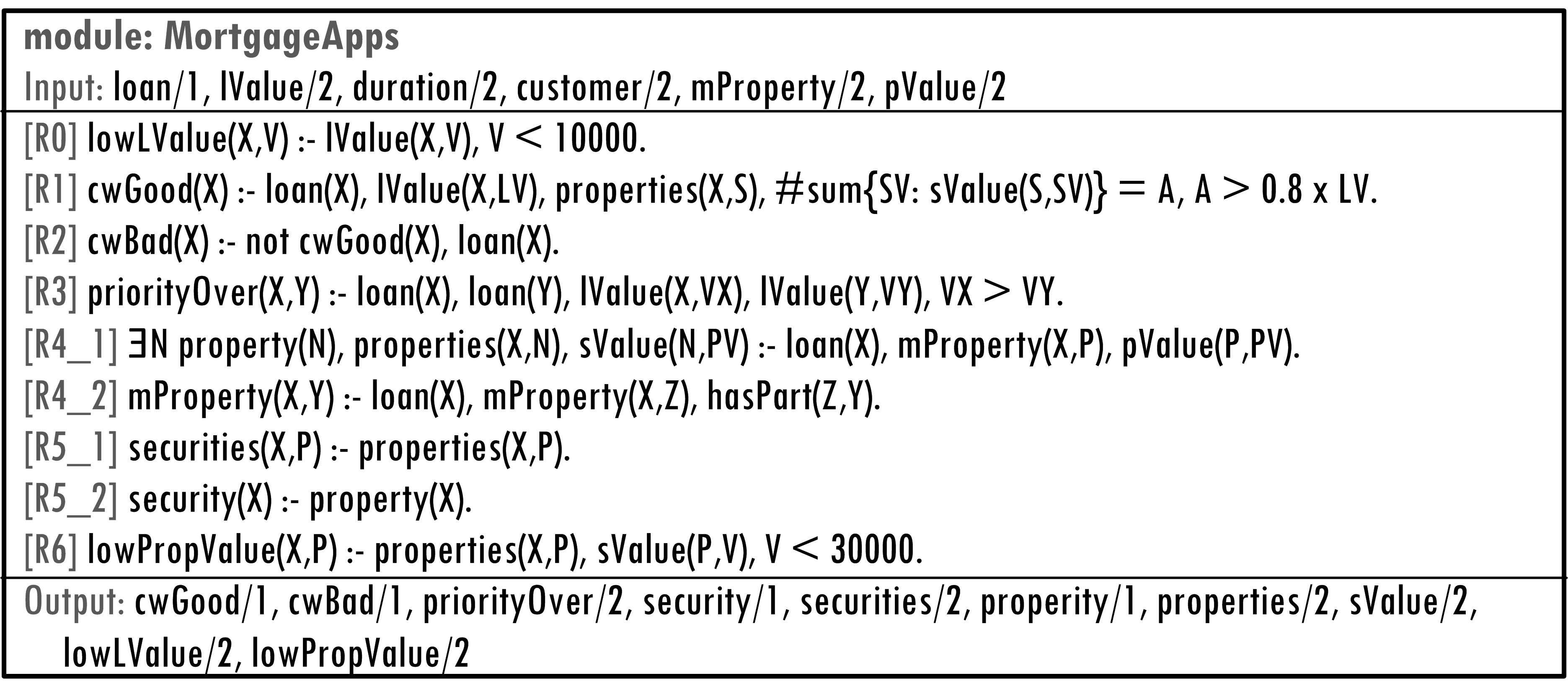}
      	\caption{Visual summary of module \code{MortgageApps} as described in \eref{ex:rule} and \ref{ex:module} comprising three compartments: the \emph{Input} and \emph{Output} compartments containing input and output declarations of predicate names and their arity respectively and the \emph{rules} compartment containing \dlpm~rules with rule identifiers in brackets.}\label{fig:MortgageApps}
      \end{figure}
\end{example}

Besides structure, a rule module exhibits behavior when executed. We regard rule module \emph{behavior} as observable effects, i.e., derived facts, when applying the set of rules in a module to a dataset containing facts for its input predicates. Multiple facts for each predicate may be provided. A data set providing facts for all input predicates of a module is called applicable.

\begin{definition}[Module Execution]\label{def:execution}
  Data sets are taken from a universe of data sets $D$.
  A data set $d \in D$ with schema $P_d \subseteq P$ contains extensions, i.e., sets of facts, for all $p \in P_d$.
  A data set $d$ is applicable input for module $m$ if the module's input schema is a subset of the data set's schema, i.e., if $I_m \subseteq P_d$.
  The execution of a module $m$ on applicable input data $d$ results, for each output predicate $p \in O_m$, in a set of derived facts, denoted as $p^d_m$ ($p^d_m$ is the set of derived $p$-facts resulting from executing $m$ on $d$).
\end{definition}

\begin{example}\label{ex:execution}
  Executing module \code{MortgageApps} on a dataset of two mortgage applications \code{L1} and \code{L2}, its rules are evaluated and derived facts of output predicates (behavior) are returned (see \fref{fig:execution}).

    \begin{figure}[tb]
    	\centering
    	\includegraphics[width=\textwidth]{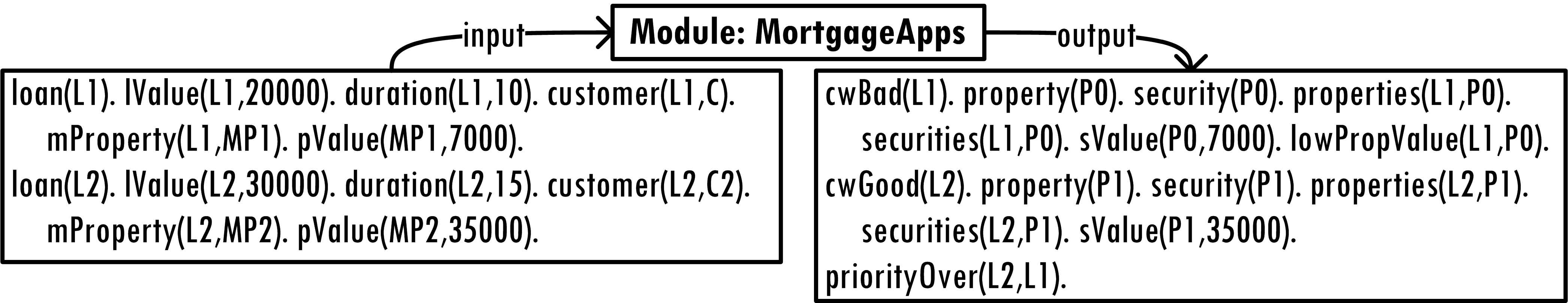}
    	\caption{Visualization of the execution of module \code{MortgageApps} on a dataset containing facts describing the two mortgage applications \code{L1} and \code{L2} and its output.}\label{fig:execution}
    \end{figure}
\end{example}

\section{Inheritance}\label{sec:inheritance}
In the following, we shortly summarize different aspects of inheritance in general. We describe the scope of our approach regarding the found aspects and subsequently define our notion of inheritance hierarchy for rule modules, rule inheritance therein, and abstract predicates and modules.

A common notion of inheritance is inheritance as incremental modification, that is, ``reusing a conceptual or physical \emph{entity} in constructing an incrementally similar one''~\cite[p.~55]{Wegner1988}. From related work we extracted various aspects and their options: All found inheritance mechanisms are transitive. Furthermore, inheritance is often discussed regarding certain \emph{foci}, i.e., signatures (schema of inputs and outputs), behavior, or implementation~\cite{Wegner1988,Pachet1992,Lang1997,Schrefl2002,Meyer1992,Kifer2017,Bertino2000}. Besides these aspects, inheritance is usually distinguished into single-inheritance and \emph{multi-inheritance}~\cite{Carbonell1980,Lang1997,Pachet1992,Schrefl2002,Wegner1988,Kifer2017}, i.e., child entities inherit from a single or multiple parent entities respectively. \emph{Direction} of inheritance is distinguished into downward~\cite{Carbonell1980,Kifer2017}, e.g., inheritance in Java, upward~\cite{Carbonell1980,Kifer2017}, e.g., an array inheriting properties from its entries, and lateral~\cite{Carbonell1980}, e.g., a motorized trike is a motorcycle except that is has three wheels.
 \emph{Granularity}~\cite{Wegner1988,Kifer2017} regards whether an inheritance mechanism is specified for groups of entities or single entities.

We restrict our investigation of rule module inheritance to the following inheritance options:
\emph{Focus} is on \emph{signature} and \emph{behavior} inheritance.
\emph{Multi-inheritance} is not investigated.
\emph{Direction} of inheritance is \emph{downward}, from parent module to child module.
\emph{Granularity} is rule modules.

\begin{definition}[Inheritance Hierarchy]
  Rule modules are arranged in an inheritance hierarchy $H \subset M \times M$ which forms a forest (i.e., a set of trees).
  We say module $m'$ inherits from $m$ if $(m',m) \in H$, with $m'$ playing the role of \emph{child} and $m$ playing the role of \emph{parent}.
\end{definition}

Rules and sets of interface predicates (input and output predicates) are propagated from parent module to its child modules. A child module is modified by introducing additional rules or interface predicates and/or by deleting inherited rules or interface predicates. Thus, rules and interface predicates can only be overridden by deletion and addition. Modifying rules implies modification of a rule module's behavior.

\begin{definition}[Inheritance by Incremental Modification]\label{def:inheritance}
  When a module $m'$ inherits from module $m$, $(m',m) \in H$, the child module $m'$ inherits the parent's rule set $R_m$ from which it may remove a set of rules $R^-_{m'} \subseteq R_m$ and to which it may add a set of rules $R^+_{m'} \subseteq R$, which results in the child module's rule set $R_{m'} \defeq R_m \cup R^+_{m'} \setminus R^-_{m'}$.
  Inheritance and incremental modification of the sets of input and output predicates are defined analogously, i.e., $I_{m'} \defeq I_m \cup I^+_{m'} \setminus I^-_{m'}$ and $O_{m'} \defeq O_m \cup O^+_{m'} \setminus O^-_{m'}$.
\end{definition}

Often discussed, in particular regarding object-orientation, are \emph{abstract entities}. For instance, an abstract method is a method for which the signature is defined but no implementation is available~\cite{Meyer1992,Wegner1988,Schrefl2002}. Methods fully defined and implemented are usually called \emph{concrete}. Analogously, we distinguish predicates in a module into concrete and abstract. We define concrete predicates as the union of predicates which are in the input interface, truth value \code{true} (represented as a nullary predicate), and predicates which contain only concrete predicates in the body of any rule having them in the rule head. Abstract predicates are defined as predicates in a module which are not concrete. In object-oriented design, a class is abstract if it contains abstract elements. Analogously, we call a rule module abstract if it contains abstract predicates and concrete otherwise. Similar to abstract classes which cannot be instantiated, abstract modules should not be applied as their behavior is incomplete. Consequently, abstract predicates and modules should always be concreted in descendant modules. Leaf modules in the module hierarchy should always be concrete.

\begin{definition}[Abstract Predicates and Modules]\label{def:abstract}
  A predicate $p$ depends on a predicate $p'$ in module $m$, denoted as $dep_m(p,p')$, if there is a rule $r$ in $R_m$ which has $p$ in the head and $p'$ in the body, i.e., $dep_m(p,p') \defeq \exists r \in R_m : p \in H_r \wedge \, p' \in B_r$. A predicate $p$ is \emph{concrete} for a module $m$ if it is nullary predicate {\tt\small true} or an input predicate or depends on some and only concrete predicates, i.e., if $(p = $ {\tt\small true}$) \vee \, (p \in I_m) \vee ((\forall p' : dep_m(p,p') \rightarrow concrete_m(p')) \wedge (\exists p' : dep_m(p,p')))$. A predicate $p \in P_m$ is \emph{abstract} for a module $m$ if it is not concrete. A module is abstract if it has an abstract predicate.
\end{definition}

From an organizational viewpoint, inheritance of rule modules can be employed for rule elicitation, definition, and organization: (a) An existing module can be adapted to a more specific setting and context by constructing a child module. Therefore, rule developers have to know which rules and interface predicates are contained in a module. To this end, they look at resolved modules, i.e., modules for which all inheritance relations have been resolved. (b) A parent module can be constructed by extracting common/similar rules and interface predicates from child modules. Abstract modules are of importance to the latter approach allowing to extract common predicates although the rules defining those predicates are different in child modules. Hierarchical organization of modules eases management, in particular maintenance, as rule redundancy can be reduced and rule reuse is promoted.

\begin{example}\label{ex:inheritance}
  Recently, our bank decided to extend its services to private loan applications. Therefore, we elicited relevant rules for private loan applications. We uncovered overlaps with module \code{MortgageApps}, in particular rules \code{R0-R3}. Rule \code{R0} overlaps with ``Private loan values must exceed 12,000'' (for now \code{RX}), \code{R1} with ``Credit worthiness of a private loan application is deemed good (predicate \code{cwGood/1}) if the value of provided securities exceeds 60 \% of the loan's value'' (for now \code{RY}), and rules \code{R2} and \code{R3} are identical. Non-overlapping are: Regarding income, private loans are reported if the customer earns less than 600 Euro per month (\code{R7}). The provided attachable income (\code{attachableIncome/1}) is calculated as 30 \% of the income earned over the loan's duration and associated with the loan application (\code{R8} -- \code{incomes/2}) as security (\code{R9} -- \code{securities/2, security/1}) allowing to reuse rule \code{R1} for deriving credit worthiness. Thus, $O_{{\tt PrivateLoanApps}} =$ $\{$\code{cwGood/1}, \code{cwBad/1}, \code{attachableIncome/1}, \code{incomes/2}, \code{priorityOver/2}, \code{lowLValue/2}, \code{sValue/2}, \code{lowIncome/2}$\}$.

  For these rules we identified the required inputs private loan application (\code{loan/1}), the intended loan value (\code{lValue/2}), the duration (\code{duration/2}), the applying customer (\code{customer/2}), and any incomes which may be used as securities (\code{income/2}). These predicates form the input predicates $I_{{\tt PrivateLoanApps}}$ which considerably overlap with $I_{{\tt MortgageApps}}$.

  A subsequent discussion with domain experts revealed that rules \code{R0}, \code{R2}, \code{R3}, and \code{RY} are actually default rules applying to all existing and future loan types. We extract the default rules and \code{MortgageApps}'s and \code{PrivateLoanApps}'s common interface predicates into a parent module \code{LoanApps}. Rule \code{RX} is renamed to \code{R0.1}, \code{R1} to \code{R1.1}, and \code{RY} to \code{R1}. Regarding interfaces we have $I_{{\tt LoanApps}} = \{$\code{loan/1}, \code{lValue/2}, \code{duration/2}, \code{customer/2}$\}$ and $O_{{\tt LoanApps}} =$ $\{$\code{cwGood/1}, \code{cwBad/1}, \code{priorityOver/2}, \code{lowLValue/1}, \code{sValue/2}, \code{securities/2}, \code{security/1}$\}$.  Since predicates \code{securities/2}, \code{security/1}, and \code{sValue/2} are not derived in \code{LoanApps} but in its child modules, module \code{LoanApps} is abstract. A visual summary of our use case including restrictions and \dlpm~representations is given in \fref{fig:rulesets}. There, rule \code{R0.1} overrides rule \code{R0} and rule \code{R1.1} rule \code{R1}, i.e., \code{$R^-_{{\tt PrivateLoanApps}}=\{$\code{R0}$\}$} and $R^+_{{\tt PrivateLoanApps}} = \{$\code{R0.1, R7, R8, R9\_1, R9\_2}$\}$. \fref{fig:privateResolved} depicts module \code{PrivateLoanApps} with inheritance resolved.

  \begin{figure}[t!]
  	\centering
  	\includegraphics[width=\textwidth]{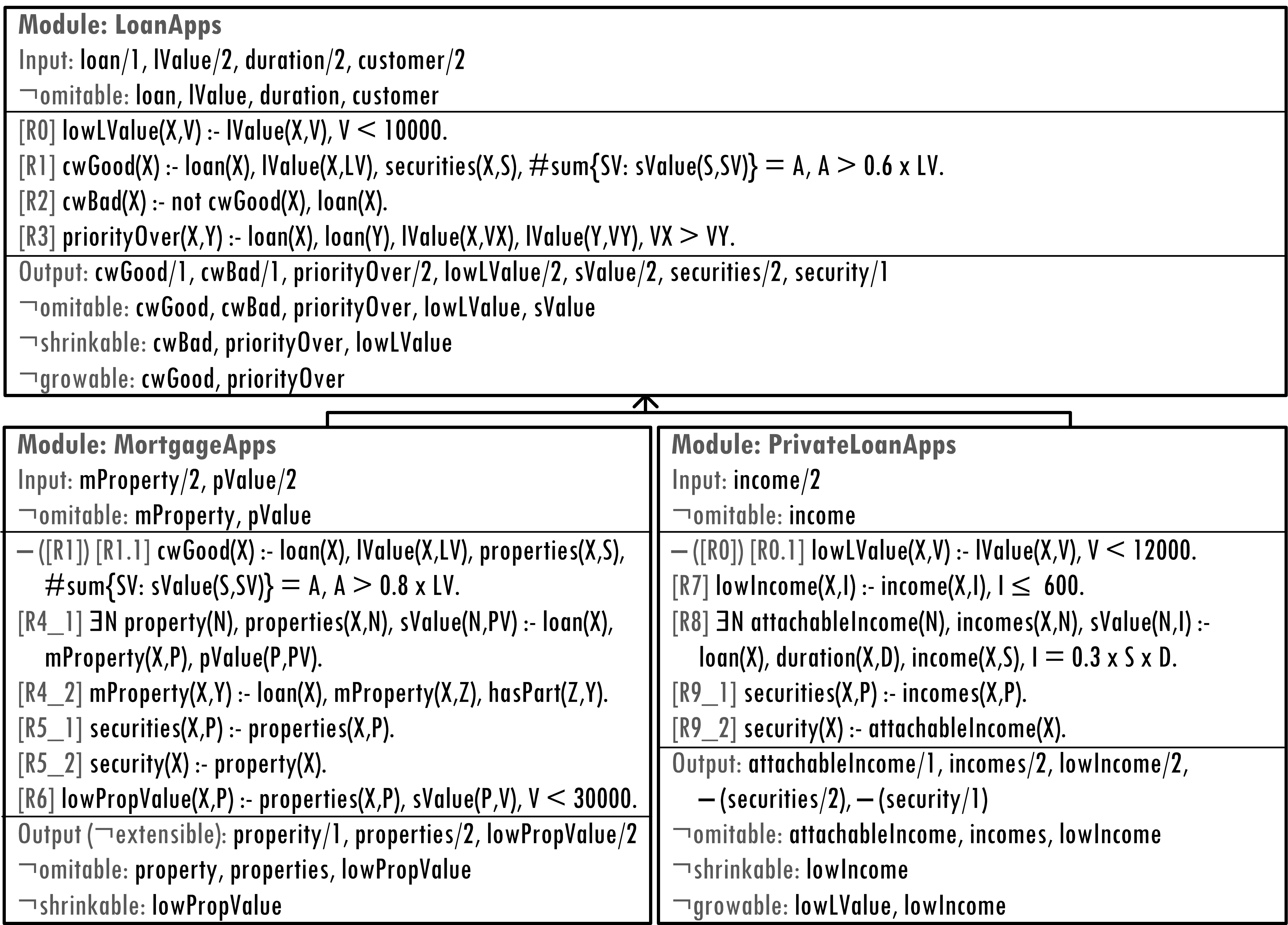}
  	\caption{Visual summary of the inheritance hierarchy of rule modules as described in Examples \ref{ex:inheritance}--\ref{ex:inheritanceMod}. Child modules \code{MortgageApps} and \code{PrivateLoanApps} (which may have their own child modules which are not depicted) incrementally modify parent module \code{LoanApps} by adding and removing (denoted by '--\code{()}') rules, input declarations, and output declarations. Input and output compartments further contain modification restrictions (denoted by $\neg$\code{omitable} and $\neg$\code{extensible}, $\neg$\code{shrinkable}, $\neg$\code{growable}) with child modules adding additional modification restrictions.}\label{fig:rulesets}
   \vspace*{\floatsep}
  	\centering
  	\includegraphics[width=.9\textwidth]{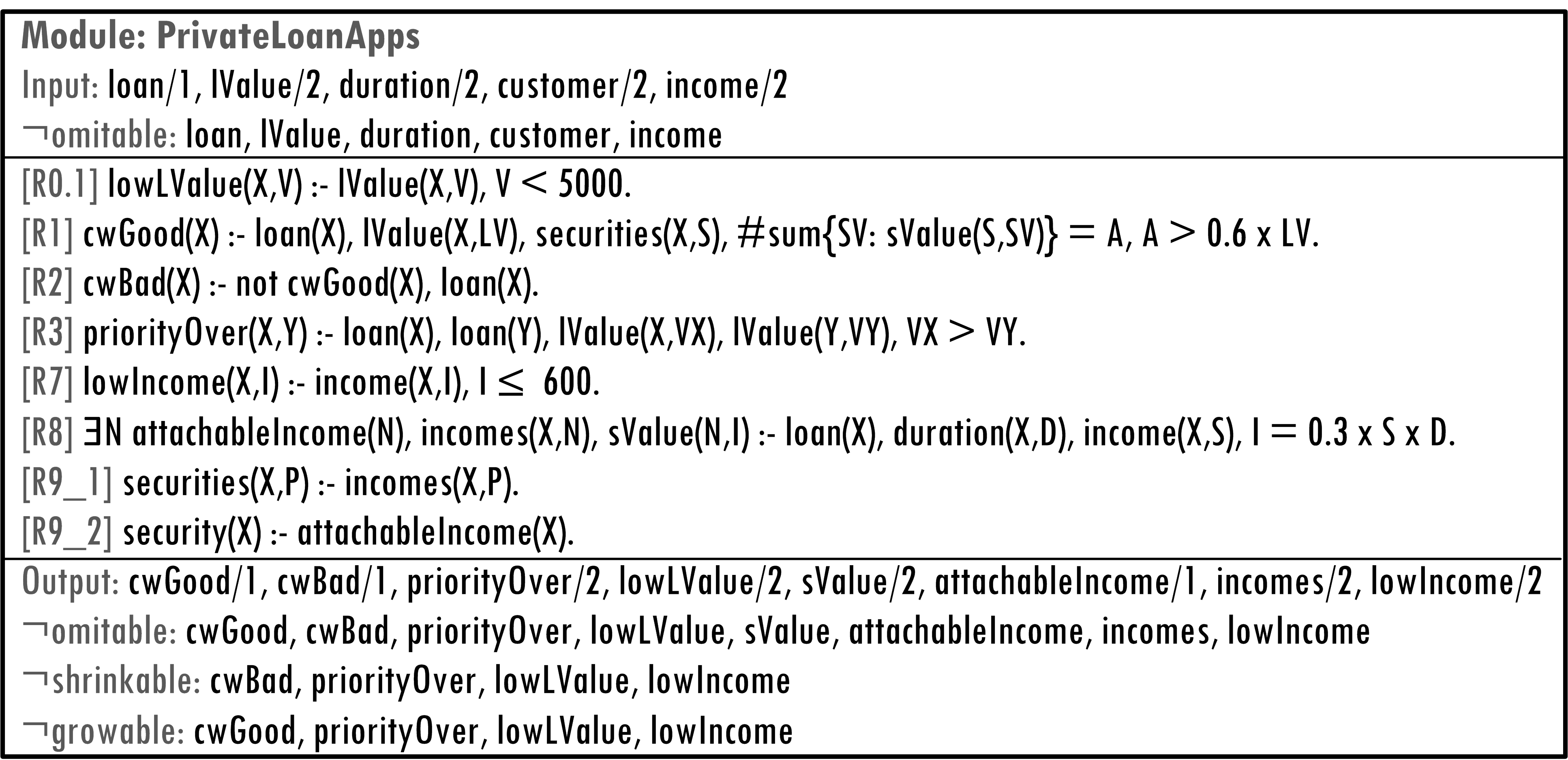}
  	\caption{Visual summary of module PrivateLoanApps with inheritance of rules, input declarations, and output declarations resolved according to \dref{def:inheritance} and inheritance of modification restrictions resolved according to \dref{def:inheritanceRestr}.}\label{fig:privateResolved}
  \end{figure}
\end{example}

\FloatBarrier
\section{Modification Restrictions}\label{sec:restrictions}
Of particular interest to this paper are modification restrictions constraining the allowed modifications by child modules. We identified the following kinds of modifications: extension, elimination, and redefinition. Extension adds features to child entities~\cite{Wegner1988,Schrefl2002,Schrefl2013,Lang1997,Pachet1992,Bichler1995,Kifer2017}. The contrary modification is elimination (also called reduction) removing features~\cite{Wegner1988,Schrefl2013,Schrefl2002}. Redefinition redefines inherited features but does not eliminate features~\cite{Wegner1988,Pachet1992,Lang1997,Kifer2017}.

In this paper, in order to keep the approach simple and compact, we focus on extension and elimination which can be used to simulate redefinition. We introduce restrictions for module structure prohibiting (a) to extend interfaces and (b) to remove specific predicates from interfaces. Regarding module behavior we introduce restrictions prohibiting: (c) to extend a module's behavior and (d) to reduce a module's behavior.

\begin{definition}[Modification Restrictions]
	A module $m$ may define a set of modification restrictions $S_m$ of the following forms: {\em no\_additional\_input}, \emph{non\_omitable\_input($p$)} with $p \in I_m$, \emph{no\_additional\_output},  \emph{non\_omitable\_output($p$)} with $p \in O_m$, \emph{non\_growable($p$)} with $p \in O_m$, and \emph{non\_shrinkable($p$)} with $p \in O_m$.
\end{definition}

\subsection{Structural Modification Restrictions}
In order to regulate modification operations on module interfaces, we introduce four restrictions. The restrictions \emph{no\_additional\_input} and \emph{no\_additional\_output} prohibit the addition of predicates to the input and output interface respectively. The restrictions \emph{non\_omitable\_input} and \emph{non\_omitable\_output} prohibit to remove the specified predicate from the respective interface.

\begin{definition}[Consistent Structural Modification]
	Let module $m'$ inherit from module $m$, $(m',m) \in H$. Structural modifications in child module $m'$ are consistent with modification restrictions in parent module $m$ if the following conditions hold:
	\begin{enumerate}
		\item if $no\_additional\_input \in S_m$ then $I^+_{m'} = \emptyset$,
		\item if $non\_omitable\_input(p) \in S_m$ then $p \notin I^-_{m'}$,
		\item if $no\_additional\_output \in S_m$ then $O^+_{m'} = \emptyset$, and
		\item if $non\_omitable\_output(p) \in S_m$ then $p \notin O^-_{m'}$.
	\end{enumerate}
\end{definition}

\begin{example}\label{ex:structuralMod}
  During the rule elicitation process our domain experts determined several modification restrictions regarding module \code{LoanApps}: Any application for a specific loan type must contain at least the same information as an application for a generic loan. Consequently, we define: $non\_omitable\_input($\code{loan}$),$ $non\_omitable\_input($\code{lValue}$),$ $non\_omitable\_input($\code{duration}$),$ $non\_omitable\_input($\code{customer}$)$. Furthermore, any loan application module must output at least the predicates for credit worthiness, priority, low loan values, and security values. Output securities may be replaced by more specific forms of outputs. Thus, we define \code{cwGood}, \code{cwBad}, \code{priorityOver}, \code{lowLValue}, and \code{sValue} as $non\_omitable\_output$. In \fref{fig:rulesets} these restrictions are listed under $\neg omitable$ in the respective interface.

  Module \code{MortgageApps} finalizes the output for any rule module inheriting from it, i.e., such a module may employ more input predicates for determining securities and their value but may not output additional predicates. Therefore, we define $no\_additional\_output$ for module \code{MortgageApps}. In \fref{fig:rulesets} we denoted this as $\neg extensible$ written next to the output of module \code{MortgageApps}.
\end{example}

From an organizational perspective, determining structural restrictions is part of rule elicitation, definition, and organization. Conformance checking of structural restrictions is necessary whenever an interface of a rule module is changed, a new rule module is added to the module hierarchy, or any structural restrictions of an existing module are modified. In the latter case, conformance of any child module of the modified module with the modified module's restrictions needs to be checked as well. A prerequisite for structural conformance checks is the identification of performed modifications. This is achieved by simple interface comparisons, e.g., parent's input schema with child's input schema.

\subsection{Behavioral Modification Restrictions}
Modifying rules contained in a rule module influences the module's behavior with respect to derived facts of output predicates. A child module, when applied on a dataset (see Def.~\ref{def:execution}), may return for a specific predicate the same, a superset, a subset, or a subset of a superset of derived facts compared to its parent module. To regulate behavioral modifications, we introduce the restrictions non\_growable and non\_shrinkable for output predicates. The restriction \emph{non\_growable} prohibits the derivation a superset of facts for a predicate in child modules whereas \emph{non\_shrinkable} prohibits child modules from deriving a subset of the facts derived by the parent module for a predicate.

\begin{definition}[Consistent Behavioral Modification]
  Let module $m'$ inherit from module $m$, $(m',m) \in H$.
  Behavioral modifications in child module $m'$ are consistent with modification restrictions in parent module $m$, if the following conditions hold for every data set $d \in D$ which is applicable to both $m$ and $m'$, as well as for every predicate $p$ which is in the output of $m$ and $m'$:
  \begin{enumerate}
  \item if $non\_growable(p) \in S_m$ then $p^d_{m'} \subseteq p^d_m$ and
  \item if $non\_shrinkable(p) \in S_m$, then $p^d_{m'} \supseteq p^d_{m}$.
  \end{enumerate}
\end{definition}

\begin{example}\label{ex:behavioralMod}
  Regarding behavior, domain experts reported several restrictions for module \code{LoanApps}:  the basic rule for good credit worthiness (\code{R1}) is the minimum requirement, i.e., its condition may only be stricter. Thus, we define $non\_growable($\code{cwGood}$)$ meaning that a loan application not deemed credit worthy according to the rules in parent module \code{LoanApps} may not be derived as credit worthy by child modules. Since every loan application is classified either \code{cwGood/1} or \code{cwBad/1} we state $non\_shrinkable($\code{cwBad}$)$. Furthermore, the loan value in \code{R0} is the minimum threshold for loan values. Consequently, the value may only be increased when specializing module \code{LoanApps}, represented as $non\_shrinkable($\code{lowLValue}$)$. As every loan application must be prioritized, behavior regarding \code{priorityOver/2} must not change, represented as $non\_growable($\code{priorityOver}$),$ $ non\_shrinkable($\code{priorityOver}$)$.
\end{example}

From an organizational viewpoint, behavioral restrictions are determined during rule elicitation, definition, and organization. To perform \emph{behavioral conformance checks}, performed behavioral modifications are compared with behavioral modification restrictions. This check can be performed during rule module testing or during execution: (a) Before modifications to a module are disseminated and deployed, they need to be thoroughly tested, i.e., the module, any parent module, and any child modules are tested with various input data sets. In addition to traditional testing, we then employ modification detection (see below) to check conformance to defined modification restrictions. (b) We can also compare the behavior of parent and child module at runtime when the child module is executed. Any violated behavioral restrictions are reported.

\subsection{Detection of Behavioral Modifications}
To determine performed behavioral modifications we propose: (a) asking responsible rule developer(s) to state his/her performed behavioral modifications manually, and (b) to automatically detect performed behavioral modification operations employing static or dynamic detection.

We expect complete static detection (by automatic reasoning over Datalog programs) of behavioral modification operations to be undecidable due to rule dependencies as well as references and predicates within input data. Nevertheless, partial detection is feasible by comparing a child module's with its parent module's rule dependency graph. A proof that this problem is undecidable as well as an algorithm for partial static detection are beyond this paper.

We now introduce dynamic detection of behavioral modifications which can be used as part of testing or during rule execution. Therefore, a child module and its parent module are executed on the same input data and their output facts compared. For a specific parent module $m$ and child module $m'$ we select a set of data sets from $D$ applicable to both the child and the parent module. For each data set $d$, we execute both modules and compare the derived facts for concrete predicates; abstract predicates are not considered in conformance checks as their behavior is incomplete. For each predicate $p$ concrete in the parent and child module, we compare $p^{d}_{m'}$ with $p^{d}_{m}$. If $p^{d}_{m'} \subset p^{d}_{m}$ the set of output facts has shrunk, if $p^{d}_{m'} \supset p^{d}_{m}$ the set of output facts has grown, if the sets of output facts has neither shrunk nor grown and $p^{d}_{m} \neq p^d_{m'}$ it has shrunk and grown, and lastly $p^{d}_{m} = p^{d}_{m'}$ implies no changes. The overall modification in behavior for a specific predicate is the union of all detected modifications in behavior. These detected behavioral modifications must not violate specified behavioral modification restrictions. The more data sets from $D$ are employed, the more reliable the detected modification(s) in behavior is/are.

\subsection{Inheritance of Modification Restrictions}
In order to achieve transitive conformance to modification restrictions we need to introduce inheritance of modification restrictions. Basically, child rule modules must not remove any modification restrictions imposed on their ancestors.

\begin{definition}[Inheritance of Modification Restrictions]\label{def:inheritanceRestr}
  Let module $m'$ inherit directly from module $m$, $(m',m) \in H$.
  The child module's set of modification restrictions $S_{m'}$ is the union of the set of modification restrictions $S_m$ inherited from parent module $m$ and the set of modification restrictions $S^+_{m'}$ added by the child module $m'$, i.e., $S_{m'} \defeq S_m \cup S^+_{m'}$.
\end{definition}

\begin{example}\label{ex:inheritanceMod}
  Module \code{PrivateLoanApps} inherits from module \code{LoanApps}. Besides rules and interface predicates, the defined modification restrictions are inherited. Consequently, any modification restriction defined in module \code{LoanApps} must hold in module \code{PrivateLoanApps} as well. This is depicted in \fref{fig:privateResolved} where inheritance has been resolved for module \code{PrivateLoanApps}, e.g., $non\_omitable($\code{cwGood}$)$ defined in module \code{LoanApps} must also hold in module \code{PrivateLoanApps}.
\end{example}

\section{Proof-of-Concept Prototype}\label{sec:eval}
Our proof-of-concept prototype implements the presented formal definitions, structural and behavioral conformance checks, and proposed modification detections. To this end, we need meta-representations of rule modules including their interfaces, inheritance relations, modification operations, and modification restrictions. Therefore, we embedded our prototype in an environment able to generate meta representations of \dlpm~programs (e.g. Vadalog~\cite{Bellomarini2018}).

To facilitate widespread experimentation we provide a download\footnote[1]{{\scriptsize\url{http://files.dke.uni-linz.ac.at/publications/burgstaller/ODBASEPrototype.zip}}} containing a Datalog implementation of our prototype (\code{terms.datalog}) and Datalog meta-representations for the rule modules in our use case (\code{<module>Meta.datalog}). The meta-representations do not include built-ins. While Datalog code is, beyond this use-case, in general severely limited in expressive power, it allows experimenting with our prototype independently of the concrete \dlpm~engine.

Executing our prototype on meta-representations of rule modules resolves inheritance, i.e., determines rules, facts, and modification restrictions applying in each module, reports abstract predicates and modules, and detects violations of structural modification restrictions. In order to test dynamic behavioral modification detection, we provide meta-representations of the output facts resulting from executing module \code{LoanApps} and module \code{PrivateLoanApps} on the same input facts (\code{result<module>.datalog}).

Executing the prototype for our use case using DLV yields reasonable performance considering that the average times include DLV start-up and no performance optimizations have been carried out. Resolving inheritance, reporting abstract predicates and modules, detecting behavioral modification operations, and structural and behavioral conformance checking take on average 0.017 s (sd = 0.001 s) on a standard notebook (Intel Core i5-6200U, DDR4, 16 GB).

\section{Related Work}\label{sec:related}
In the following we discuss inheritance of single rules; rule sets, rule modules, and their inheritance; as well as contextual knowledge and its inheritance. Several researchers proposed inheritance of pre- and postconditions of operations (rule premises) \cite{Meyer1992,OMG2014,Schrefl2002} where modifications to conditions are restricted to strengthening or weakening. Other research regards inheritance of triggers (which can be considered event-condition-action rules) in object-oriented databases \cite{Bertino2000}, where a trigger's premises may become weaker and trigger actions may be extended. \flora~\cite{Kifer2017} combines rules and object-oriented features where methods are specified as rules and can be inherited.

\subsection{Rule Sets, Modules, and Inheritance}
Modularization is employed in many fields, e.g., software engineering or ontologies, enabling controlled and structured development of large systems~\cite{Parent2009}. Regarding knowledge, a knowledge base may either be partitioned into non-overlapping modules or relevant portions extracted into possibly overlapping modules~\cite{Parent2009}. The concrete modularization strategy depends on the use case.

Regarding ontologies, a triple containing an ontology, a query language, and a vocabulary can be considered a module, where the query language and the vocabulary serve as interface allowing interaction by querying~\cite{Konev2009}. A simple extension mechanism enabling addition of ontological statements is described~\cite{Konev2009}.

Rule modules with relational schemas as interface specifications are a simplified variant of relational transducers~\cite{Abiteboul1998}. In the original proposal, relational transducers serve as ``declarative specifications of business models, using an approach in the spirit of active databases''. Relational transducers transform sequences of input into sequences of output relations. In addition to input and output relations, a relational transducer specifies database, state, and log relations, where the log relations are the semantically relevant subset of input and output relations. Regarding inheritance, they discuss customization of relational transducers and with regard to restricted modification they discuss ``containment and equivalence of relational transducers relative to a specified log''.

Modular Web rule bases~\cite{Analyti2011} separate interfaces, i.e., predicates used and predicates defined, from the logic program, i.e., rules.
Each predicate in a module defines its reasoning mode, its scope, and its origin rule bases or rule bases it is visible to. Modular rule base extension is supported, allowing to add new rules to existing rule bases and to add new rule bases to the set of rule bases.

Inheritance of (business) rules is touched in~\cite{Burgstaller2016,Burgstaller2017c} (see contextualized knowledge below) and discussed specific to situa\-tion-condition-actions rules in~\cite{Lang1997}. Inheritance of rule sets is discussed in~\cite{Pachet1992}. Lang~\cite{Lang1997} identifies the rule of origin as prerequisite, i.e., a feature may only be defined in a single point. Moreover, he utilizes the abstract parent class rule. Situations, conditions, and actions may be specialized if the occurrence of a situation implies occurrence of its child situations and rule conditions are only weakened. Based on these conditions modification operations are proposed~\cite{Lang1997}: extension denotes addition of rules or redefining rules to fire more often, refinement denotes concreting abstract rules, i.e., concreting rule-triggering interval classes to event classes, and redefinition is constrained to specialized actions and events as well as weakened conditions. Pachet~\cite{Pachet1992} describes rule set inheritance as inheriting all rules from parent rule sets and allows for unconstrained redefinition of rules.

A related field are business rule management systems (BRMSs) like IBM's JRules, JBoss Drools, FICO Blaze Advisor, or Oracle Business Rules, which organize business rules into rule sets. JRules supports inheritance of rules and rule sets where rules may be overridden. Drools and Blaze both support inheritance of rule conditions. Oracle does not report support for rule (set) inheritance.

Many of the above approaches support rule sets (\cite{Parent2009,Konev2009,Burgstaller2016,Burgstaller2017c,Lang1997,Pachet1992,Kifer2017}, BRMSs) but only  Abiteboul et al.~\cite{Abiteboul1998}, Analyti et al.~\cite{Analyti2011}, and Konev et al.~\cite{Konev2009} describe modules with explicit definition of input and output interfaces as supported by our approach. Rule set inheritance is supported by quite a few approaches, where some allow no or only predefined modification types (\cite{Abiteboul1998},~\cite{Konev2009},~\cite{Lang1997},~\cite{Burgstaller2017c}, Drools, Blaze) and others allow arbitrary modifications (\cite{Burgstaller2016},~\cite{Pachet1992}, JRules). Nevertheless, none of them provides any means to assert fine-grained control over the types of modifications to inherited rule sets as our approach does with the presented modification restrictions. By these modification restrictions, our approach allows, unlike the ones discussed above, to flexibly adapt the inheritance mechanism to the specific needs at hand.

\subsection{Contextualized Knowledge Representation and Inheritance}
A related field are contextual knowledge and its inheritance. Contextualized Knowledge repositories (CKR)~\cite{Serafini2012} for the Semantic Web organize ontological concepts employing hierarchically ordered contexts. A context is defined by a set of values for various but fixed context dimensions. Allowed dimension values form a subsumption hierarchy from which the hierarchy of contexts is derived. Concepts propagate along this hierarchy from general contexts to more specific contexts. Concepts may assume different meanings in different contexts. A similar idea was proposed for knowledge organization in CYC~\cite{Lenat1998} where it is explicitly allowed to contradict or override inherited knowledge. McCarthy~\cite{McCarthy1993} views contexts as generalizing collections of assumptions where a child context must have at least the same assumptions as its parent context. Knowledge from child contexts must be translatable into meaningful knowledge in the parent context, i.e., a root context would contain all knowledge in decontextualized form.

Building on these approaches, we introduced a contextualized (business) rule repository \cite{Burgstaller2016} allowing to organize rule sets into multi-dimensional context hierarchies. Regarding inheritance, we differentiate additive and most-specific inheritance of rule sets. With the former inheritance mechanism all rules of a parent rule set apply in its child rule sets as well. The latter approach allows redefinition of inherited rules. Previous work on contextualized rule repositories for the Semantic Web \cite{Burgstaller2017c} employed an additive inheritance semantics only.

Since contexts can be considered modules, all of the above approaches support modules. Nevertheless, none of them explicitly defines clear input and output interfaces for modules as our approach does. Regarding module inheritance, all of the above approaches support inheritance of knowledge, some~\cite{Lenat1998,Burgstaller2016} allow redefinition of knowledge whereas others do not~\cite{McCarthy1993,Serafini2012,Burgstaller2017c}. Compared to our approach, none of the above approaches supports modification restrictions, i.e., one cannot control modifications to the knowledge in a module.

\section{Conclusion}\label{sec:conclusion}
We investigated inheritance of rule modules to foster reuse of rules, simplify adaptation, and ease maintenance. Therefore, we introduced rule modules and proposed a formal inheritance mechanism. We presented modification restrictions regulating modifications with corresponding conformance checks as mechanisms for keeping child modules aligned with parent modules.

In ongoing and future work we investigate:
\begin{itemize}
\item extending our proposed approach to multi-inheritance and integrating it into  \emph{context-aware business rule management}~\cite{Burgstaller2017} and Vadalog~\cite{Bellomarini2018}.
\item rule modules as part of derivation chains or module networks in \emph{big data wrangling}~\cite{Konstantinou2017,Furche2016} and \emph{knowledge graph management}~\cite{Bellomarini2017a}. There, rule set adaptation is necessary to cope with the variety of relevant subject domains and the variety of integrated sources of data and knowledge. One such scenario that we want to apply our approach to is building knowledge graphs on huge elections in the area of computational social choice \cite{Csar2017}, another one is in data extraction, where inheritance hierarchies are essential for managing diverse ontologies \cite{Furche2014,Michels2017}. Furthermore, the clean interfaces of our rule modules support derivation chains and networks.
\end{itemize}

\bibliographystyle{splncs04}

\end{document}